\documentclass[aps,prb,preprint,showpacs,showkeys,floatfix]{revtex4}

\newcommand{\jwj}[1]{\textcolor{blue}{#1}}

\usepackage{graphicx,color}
\graphicspath{{figs/}}
\begin{document}
\title{Size-Sensitive Young's modulus of Kinked Silicon Nanowires}
\author{Jin-Wu Jiang}
    \altaffiliation{Electronic address: jinwu.jiang@uni-weimar.de}
    \affiliation{Institute of Structural Mechanics, Bauhaus-University Weimar, Marienstr. 15, D-99423 Weimar, Germany}
\author{Jun-Hua Zhao}
    \altaffiliation{Electronic address: junhua.zhao@uni-weimar.de}
    \affiliation{Institute of Structural Mechanics, Bauhaus-University Weimar, Marienstr. 15, D-99423 Weimar, Germany}
\author{Timon Rabczuk}
    \altaffiliation{Electronic address: timon.rabczuk@uni-weimar.de}
    \affiliation{Institute of Structural Mechanics, Bauhaus-University Weimar, Marienstr. 15, D-99423 Weimar, Germany}

\date{\today}
\begin{abstract}
We perform both classical molecular dynamics simulations and beam model calculations to investigate the Young's modulus of kinked silicon nanowires (KSiNWs). The Young's modulus is found to be highly sensitive to the arm length of the kink and is essentially inversely proportional to the arm length. The mechanism underlying the size dependence is found to be the interplay between the kink angle potential and the arm length potential, where we obtain an analytic relationship between the Young's modulus and the arm length of the KSiNW. Our results provide insight into the application of this novel building block in nanomechanical devices.
\end{abstract}

\maketitle

\section{introduction}
In 2009, a new type of building block - kinked silicon nanowire (KSiNW) - was synthesized by Tian {\it et al.} in Lieber's group at Harvard University\cite{TianB}. The growth direction of the KSiNW changes from $<112>_{\rm arm}$ to $<110>_{\rm joint}$ to $<112>_{\rm arm}$ at the kink. Particularly, the researchers are able to manipulate the arm length of the kink by controlling the growth time. Great experimental efforts have been devoted to investigating the growth mechanism of the kinks in silicon nanowires since then\cite{ChenH,YanC,KimJ,PevznerA,ShinN,MusinIR}. Besides these experimental works, Schwarz and Tersoff have proposed a theoretical model to interpret the growth mechanism of the KSiNW. In their model, the kink formation is due to the interplay of three basic processes: facet growth, droplet statics, and the introduction of new facets\cite{SchwarzKW,SchwarzKWprl}. Stimulated by KSiNWs, several groups have synthesized kinks in other nanowires, such as ${\rm In_{2}O_{3}}$ multikinked nanowires\cite{ShenG}, kinked germanium nanowires\cite{KimJH}, germanium–silicon axial heterostructure with kinks\cite{DayehSA}, and kinked ZnO nanowires\cite{LiS}. While existing works for KSiNW mainly concentrate on its growth mechanism, the study of elastic properties like Young's modulus is also important for its application in nanomechanical devices.

The bulk silicon is an anisotropic mechanical material with the value of Young's modulus distributed roughly between 110~{GPa} and 180~{GPa} depending on the lattice direction.\cite{McskiminHJ,WortmanJJ,KimKY} Hopcroft {\it et.al} have shown that the Young's modulus in bulk silicon along $<110>$ direction is about $45\%$ higher than that in the $<100>$ lattice direction.\cite{HopcroftMA} In nano materials, due to large surface to volume ratio, the size effect\cite{LieberCM} and nanodefects\cite{PugnoNM} have been found to be important for the mechanical properties such as the Young's modulus. For example, the AFM-measured Young’s modulus ranged from 93 to 250 GPa depending on the nanowire diameter.\cite{Tabib-AzarM}

In this paper, we apply both molecular dynamics (MD) simulations and finite element method to study the size-dependence of the Young's modulus in the KSiNW. We find that the Young's modulus is sensitive to its arm length; specifically, it decreases fast with increasing arm length. The underlying mechanism for this size-dependence is disclosed to be the competition between kink angle potential and the arm length potential.

\section{lattice constraint on the diameter of the KSiNW}
\begin{figure}[htpb]
  \begin{center}
    \scalebox{1.0}[1.0]{\includegraphics[width=\textwidth]{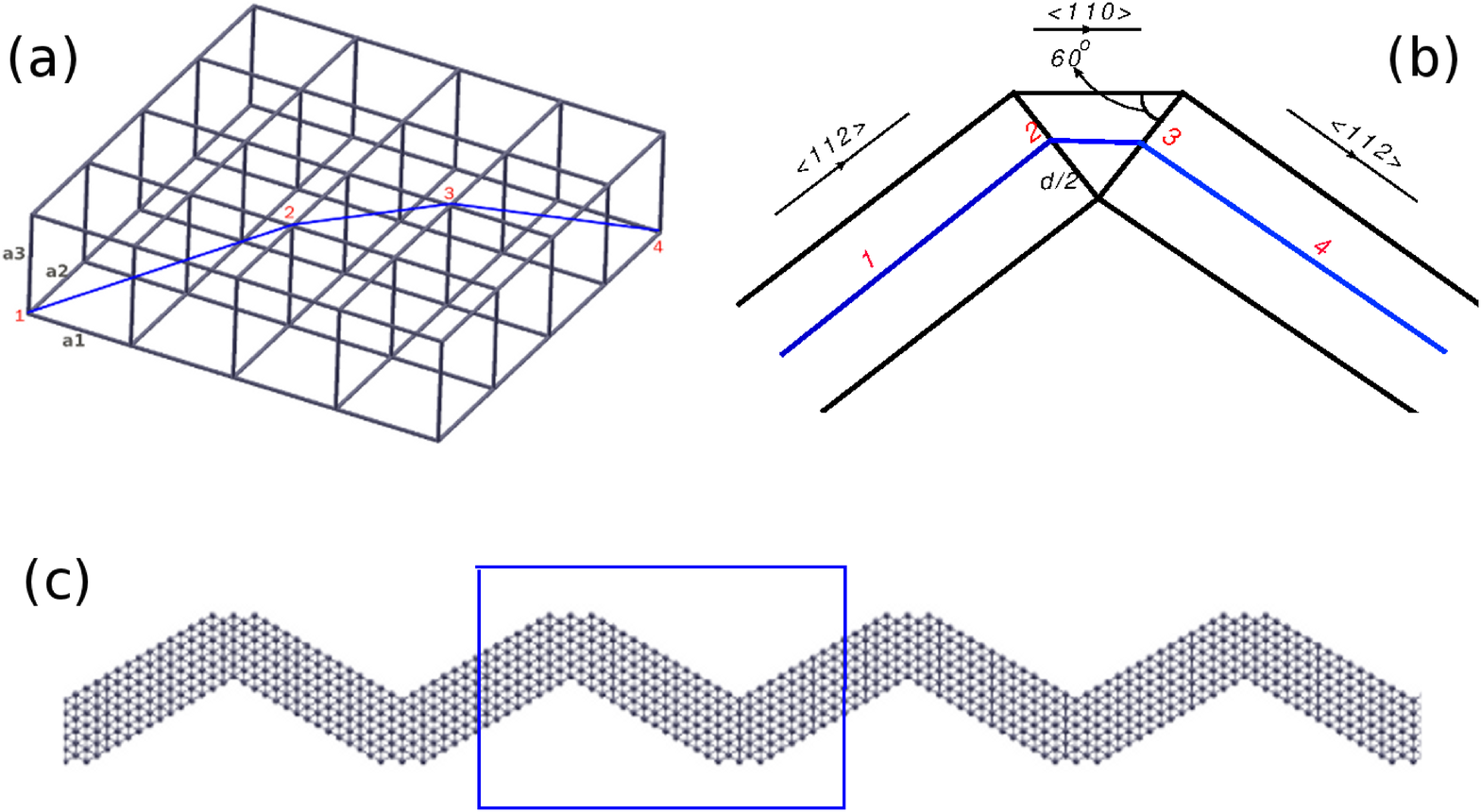}}
  \end{center}
  \caption{(Color online) The configuration of a KSiNW. (a) shows the growth axis from $<112>_{\rm arm}$ to $<110>_{\rm joint}$ to $<112>_{\rm arm}$. (b). A schematic geometry, disclosing a lattice constraint on the diameter $d$, i.e $d/2=r_{23}$. (c). Configuration of a real KSiNW with structure parameters $(n_{\rm arm}, n_{\rm joint}) = (2, 1)$. The rectangular box (blue online) highlights a unit cell of the KSiNW.}
  \label{fig_cfg_ksinw}
\end{figure}

Fig.~\ref{fig_cfg_ksinw} displays the configuration of a KSiNW. Panel (a) shows the growth axis of the KSiNW in a cubic lattice. $\vec{a}_{1}$, $\vec{a}_{2}$, and $\vec{a}_{3}$ are the three primitive vectors in the cubic lattice, with lattice constant $a=5.43$~{\AA}. Lattice vectors $\vec{R}_{12}=(2,1,1)$ and \jwj{ $\vec{R}_{34}=(1,2,-1)$} form two arms of the kink, while $\vec{R}_{23}=(1,1,0)$ forms the joint of the kink. Geometrically, the KSiNW is denoted by a pair of integers $(n_{\rm arm}, n_{\rm joint})$. The arm length is $b_{0}=n_{\rm arm}R_{12}$ and the joint length is $n_{\rm joint}R_{23}$.

From the schematic view shown in panel (b), the diameter of a KSiNW is restricted to some discrete values as determined by the lattice vector $\vec{R}_{23}$, i.e \jwj{ $d=2 r_{23}=2 n_{\rm joint} R_{23}=1.54n_{\rm joint}$~{nm}}. This lattice constraint will have direct result on the configuration of the kink. In the experiment, the diameter $d$ is pre-defined during the growth of the KSiNW\cite{TianB}, which will likely deviate from the above discrete values. \jwj{There are two possible consequences for the KSiNW to accommodate the pre-defined diameter. Firstly, the diameter of the joint part may become more difficult to be controlled in the experiment, if the diameter of the arm deviates from the discrete values. As a result, there will be a discontinuity at the kink between $<112>_{\rm arm}$ and $<110>_{\rm joint}$ growth directions. This discontinuity helps to break the lattice constraint. The obvious discontinuity in Ref.~\onlinecite{TianB} may relate to this consequence. Another possible way to break the lattice constraint is that the axial length of the joint becomes more difficult to be controlled, when the diameter of the arm deviates form the discrete values. As a result, the joint is not a perfect regular triangle anymore. Instead, the joint turns into a echelon-like structure, which is clearly shown in the Fig.2(c) of Ref.~\onlinecite{TianB}.} In KSiNW samples with diameter down to a few nanometers, the lattice constraint will play a more important role. Panel (c) shows four unit cells of a KSiNW with 1.54~{nm} in diameter and 2.66~{nm} in arm length, i.e $(n_{\rm arm}, n_{\rm joint})=(2,1)$. The unit cell is highlighted by a rectangular box.

\section{results and discussion}
\subsection{Results from MD simulation and finite element method}
In our MD simulation, the interaction between Si atoms is modeled by the widely used Stillinger-Weber empirical potential\cite{StillingerF}. The structure of the KSiNW is first relaxed without applying strain, yielding the minimum total energy $E_{0}$. After applying strain $\epsilon$ to the system, the tensile structure is optimized with left and right ends fixed, which results in a total energy $E(\epsilon)$. The axial Young's modulus is then calculated from the second derivation of the strain energy density $(E(\epsilon)-E_{0})/V$ with respective to the strain $\epsilon$ in $[0,0.01]$; $V$ is the volume. \jwj{It should be noted that the obtained Young's modulus is that of the overall kinked system (i.e. parallel to the unit cell in Fig.~\ref{fig_cfg_ksinw}~(c)), and not the Young’s modulus in the axial direction in each part of the nanowire.}

\begin{figure}[htpb]
  \begin{center}
    \scalebox{0.8}[0.8]{\includegraphics[width=\textwidth]{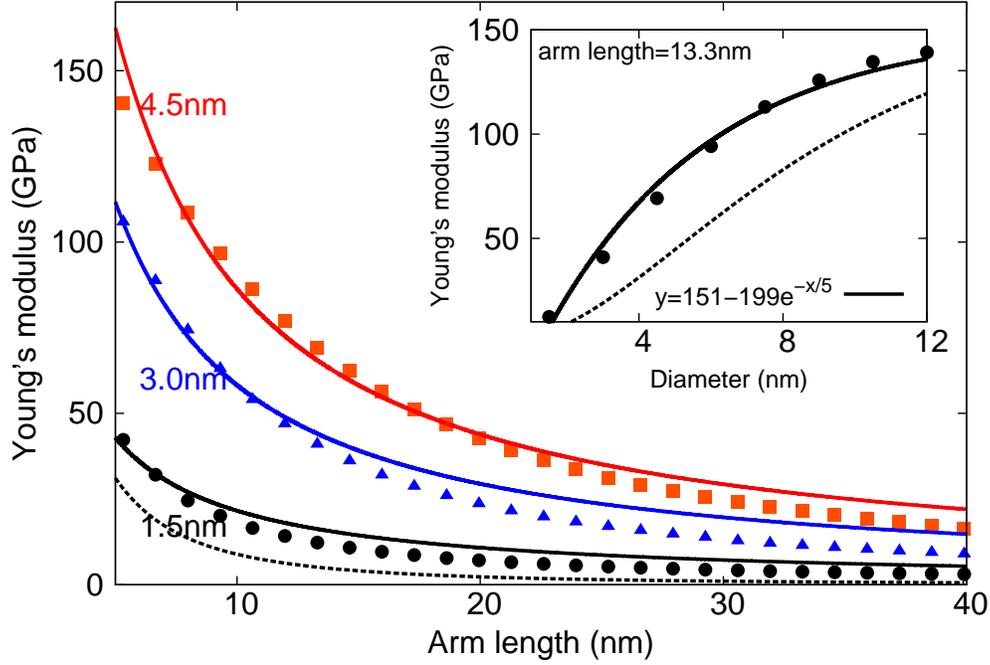}}
  \end{center}
  \caption{(Color online) The size effect on the axial Young's modulus. The MD simulation results for the Young's modulus versus arm length in KSiNW with diameters 1.54~{nm} (black points), 3.07~{nm} (blue triangulars), and 4.61~{nm} (red squares). Simulation results are compared with the valence force field model (solid lines) and the elastic beam model (dashed lines). Inset: the Young's modulus versus diameter for KSiNW with arm length 13.30~{nm}, where calculation results (points) are fitted to an exponential function (solid line) and compared with the elastic beam model (dashed line).}
  \label{fig_young}
\end{figure}

A sensitive size-dependence is observed for the Young's modulus as shown in Fig.~\ref{fig_young} for three sets of KSiNWs with diameters \jwj{1.54, 3.07, and 4.61}~{nm}. Only one unit cell is considered here, and we found that the Young's modulus does not depend on the number of unit cells in the KSiNW. The Young's modulus decreases with increasing arm length and can be fitted to an analytic function (solid lines) $y=c_{1}/(b_{0}+c_{2}/b_{0})$ which can be well explained by the valence force field model as will be shown later. The two coefficients $(c_{1}, c_{2})$ are (216.5, 0.23), (591.1, 1.45), and (881.7, 2.12) for these three sets of KSiNWs. It shows that the Young's modulus can be very small for KSiNW with very long arm length. This prediction can be readily verified experimentally, considering the arm length of KSiNW samples can be successfully controlled in the laboratory\cite{TianB}.

The inset of Fig.~\ref{fig_young} shows the diameter dependence for the Young's modulus in KSiNWs with arm length 13.3~{nm}. The Young's modulus increases exponentially and saturates to a constant value of 151.0~{GPa}. Similar diameter dependence was observed in straight silicon nanowires.\cite{RuddRE} Using the Stillinger-Weber potential, we have calculated the Young's modulus in a bulk silicon to be 159.0~{GPa} in $<110>$ direction and 132.0~{GPa} in $<112>$ direction. Both values fall in the experimental range of the Young's modulus in bulk silicon.\cite{McskiminHJ,WortmanJJ,KimKY} The saturation value of 151.0~{GPa} in large diameter limit is sandwiched between the Young's modulus in bulk silicon along $<110>$ and $<112>$ directions. This result can be understood from the fact that the KSiNW is constructed by $<110>$ (the arm) and $<112>$ (the kink) silicon nanowires.

\begin{figure}[htpb]
  \begin{center}
    \scalebox{1.2}[1.2]{\includegraphics[width=8cm]{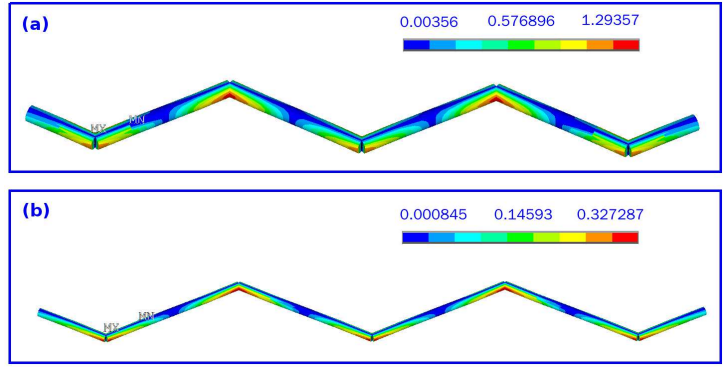}}
  \end{center}
  \caption{(Color online) Von Mises stress \jwj{(in unit of $10^{10}$~Pa)} distribution in KSiNWs from the elastic beam model calculation. The applied tensile displacement is the same in (a) for the KSiNW with arm length 13.30~{nm} and diameter 1.54~{nm}, and (b) for the KSiNW with arm length 26.60~{nm} and diameter 1.54~{nm}. Note the stress concentration at kinks. \jwj{Attention that the longer nanowire in (b) has been scaled by an overall factor of 0.5.}}
  \label{fig_beam}
\end{figure}

We also calculate the Young's modulus of KSiNWs from the continuum theory based on the finite element method. We use the Euler-Bernoulli beam model\cite{TimoshenkoS,LiH,ZhaoJH} in our calculation. Finite element calculations are performed using the commercial ANSYS 12.0 package with 2-node BEAM188 element. The only input parameters required by this package is the Young's modulus and Poisson ratio for bulk silicon from the above calculation based on the Stillinger-Weber empirical potential. Fig.~\ref{fig_beam} shows the Von Mises stress distribution from the elastic beam model for two KSiNWs with arm length 13.30~{nm} and diameter 1.54~{nm} in (a), and arm length 26.60~{nm} and diameter 1.54~{nm} in (b). The applied tensile displacements are the same in both structures. The stress in the thinner KSiNW in (b) is roughly one quarter of that in the KSiNW in (a), because the Young's modulus and the strain in the thinner KSiNW in (b) are both about half of that in the KSiNW in (a).

\jwj{We note that the Euler-Bernoulli beam model neglects the shear strain, while this shear strain is considered in the Timoshenko beam model. We have also tested the Timoshenko beam model, and a similar size dependence of the Young's modulus is obtained while the value is overall smaller. An interesting phenomenon in Fig.~\ref{fig_beam} is the distinct stress concentration at kinks. The stress concentration may lead to the generation of a fracture around the kink in the KSiNW under large tension. This stress concentration phenomenon also affects the accuracy of the predicted Young's modulus. An important parameter in such calculation is the stress concentration ratio (SCT), i.e the ratio of the stress concentration length to the total length. The SCT is less than 30\% in both KSiNWs shown in Fig.~\ref{fig_beam}. This is important. It is because the Saint-Venant principle says that it is crucial for the SCT to be less than 30\% for an accurate prediction of the Young's modulus.\cite{LoveAEH,ZhaoJH2013} According to the Saint-Venant principle,\cite{LoveAEH} a more accurate Young's modulus can be obtained from a smaller SCT, because the stress is distributed uniformly in more areas within the system in case of smaller SCT. From the comparison of these two KSiNWs in Fig.~\ref{fig_beam}, the SCT decreases with increasing arm length. As a result, the Young's modulus from the finite element method should be more accurate for KSiNWs of longer arm length. Indeed, Fig.~\ref{fig_young} shows a better agreement between the Young's modulus from the finite element method and the molecular mechanics approach for KSiNWs of longer arm length.}

Results for the set of KSiNWs of diameter 1.54~{nm} are shown by the dashed line in Fig.~\ref{fig_young}. The overall tendency from the finite element calculation agrees well with the MD simulation results. The Young's modulus decreases with increasing arm length. It should be noted that the finite element results always lie bellow the MD simulation results, which may probably due to the lose of part of the angle bending interactions in the beam model and also surface effects. As a result, the beam model actually mimics a system that is softer than the real material. The softness property of the beam model has been discussed in more detail in a recent work by Zhao {\it et.al}.\cite{ZhaoJH2012} The deviation between the finite element results and the MD simulations is more pronounced for KSiNWs with smaller arm length, as the Euler-Bernoulli beam model is more accurate for a thinner rod structure with higher length to diameter ratio. There will be some systematic errors in the Euler-Bernoulli beam model if the simulated is very thick and/or short, since the higher order terms associated with the shearing deformation are ignored in this model. The surface induced diameter dependence of the Young's modulus from the finite element method is also shown in the inset of Fig.~\ref{fig_young}. Again, the general tendency from finite element method agrees with that from the MD simulation, and the elastic method also gives a lower value for the Young's modulus.

\jwj{We note that it is a usual technique to use the bulk Young's modulus of silicon as input for finite element modeling of mechanical properties of nanomaterials or nanostructures.\cite{WangG,LiXF} Yet, there is some uncertainty in doing so, since the nanowires are anisotropic in atomic-scales. A different Young's modulus value may lead to an overall shift for the whole curve in Fig.~\ref{fig_young}, while the size dependence of the Young's modulus is kept unchanged. As a result, the size dependence of the Young's modulus from the finite element modeling still agrees with that from the molecular mechanics method, although the input Young's modulus for the finite element modeling varies. This size dependence is more important in the present work. In this sense, it is appropriate to use the bulk properties of silicon for the finite element modeling of the kinked silicon nanowires.}

\subsection{Analysis based on valence force field model}
\begin{figure}[htpb]
  \begin{center}
    \scalebox{1.2}[1.2]{\includegraphics[width=8cm]{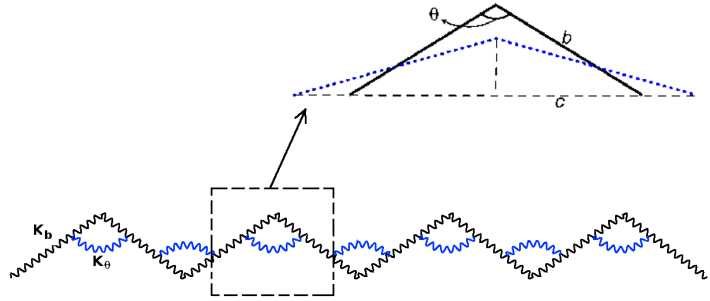}}
  \end{center}
  \caption{(Color online) The KSiNW is simplified by a valence force field model with spring constant $k_{b}$ and angle bending constant $k_{\theta}$. A single kink (inside the box) is highlighted for its equilibrium structure without strain (solid, black online) and under mechanical strain (dotted, blue online). Geometrical variables are related to each other by $c=b\sin\frac{\theta}{2}$.}
  \label{fig_cfg_kink}
\end{figure}

The above size-dependence for the axial Young's modulus can be understood in a valence force field model. Considering its zigzag configuration, the KSiNW can be simplified by a series of springs representing its arm length and kink angle as shown in Fig.~\ref{fig_cfg_kink}. The arm-spring has a force constant of $k_{b}$, while the kink-spring has a force constant of $k_{\theta}$. The potential of these two springs give the most important interaction for such a zigzag structure. Due to the translation invariance, we only need to consider the interaction within a single kink in the KSiNW. The kink angle bending ($V_{\theta}$) and the arm length stretching ($V_{b}$) potentials are\cite{ZhouX,ShenL,ChangT,JiangJW2006,JiangJW2008,JiangJW2008prb},
\begin{eqnarray}
V_{\rm tot} = V_{\theta}+V_{b} = \frac{k_{\theta}}{2}\left(\cos\theta-\cos\theta_{0}\right)^{2}+\frac{k_{b}}{2}\left(b-b_{0}\right)^{2},
\label{eq_Etot}
\end{eqnarray}
where $\theta$ is the kink angle and $b$ is the arm length. Variables with subscript 0 correspond to their values at equilibrium configuration without strain. For uniaxial strain applied in the horizontal direction, Fig.~\ref{fig_cfg_kink} shows a relationship among geometrical variables
\begin{eqnarray}
c=b\sin\frac{\theta}{2}.
\end{eqnarray}
There is only one independent degree of freedom in the tensile KSiNW. We choose the kink angle $\theta$ as the free variable. From the energy minimum condition $\partial E/\partial \theta=0$, we have
\begin{eqnarray}
&&k_{\theta}\left(\cos\theta-\cos\theta_{0}\right)\sin\theta\sin^{3}\frac{\theta}{2}\nonumber\\
&+&\frac{1}{2}k_{b}c\left(c-b_{0}\sin\frac{\theta}{2}\right)\cos\frac{\theta}{2} = 0.
\label{eq_EMC}
\end{eqnarray}

The solution of Eq.~(\ref{eq_EMC}) yields the equilibrium structure of the KSiNW under mechanical strain $\epsilon$. For small strain, the kink angle is also small, so it only slightly deviates from its equilibrium value, i.e $\theta=\theta_{0}+\delta_{\theta}$. Applying standard perturbation theory to Eq.~(\ref{eq_EMC}) up to $(\delta_{\theta})^{2}$, we get
\begin{eqnarray}
\alpha \delta_{\theta}^{2}+\beta \delta_{\theta}+\gamma = 0,
\end{eqnarray}
where the three coefficients are:
\begin{eqnarray}
\left\{
\begin{array}{l}
\alpha=k_{b}c\left(\frac{\sqrt{3}}{8}b_{0}-\frac{1}{16}c\right),\\
\beta=\left[k_{b}c\left(\frac{1}{4}b_{0}-\frac{\sqrt{3}}{4}c\right)-\frac{9\sqrt{3}}{16}k_{\theta}\right],\\
\gamma=k_{b}c\left(\frac{1}{2}c-\frac{\sqrt{3}}{4}b_{0}\right).
\end{array}
\right.
\label{eq_coefficient}
\end{eqnarray}
The solution of the equilibrium equation is then obtained analytically,
\begin{eqnarray}
\delta_{\theta}^{\pm}=\frac{-\beta\pm\sqrt{\beta^{2}-4\alpha\gamma}}{2\alpha}.
\label{eq_solution}
\end{eqnarray}

It can be shown that the elastic properties are dominated by the first-order term in the Taylor expansion of the angle variation $\delta_{\theta}$ in terms of $\epsilon$. We thus expand the three coefficients in Eq.~(\ref{eq_coefficient}) to the Taylor series of $\epsilon$:
\begin{eqnarray}
\left\{
\begin{array}{l}
\alpha \approx \alpha_{0}+\alpha_{1}\epsilon,\\
\beta  \approx \beta_{0}+\beta_{1}\epsilon,\\
\gamma \approx \gamma_{0}+\gamma_{1}\epsilon,
\end{array}
\right.
\end{eqnarray}
where
\begin{eqnarray}
\left\{
\begin{array}{l}
\alpha_{0} = \frac{9}{64}k_{b}b_{0}^{2},\\
\beta_{0}  = -\frac{\sqrt{3}}{16}k_{b}b_{0}^{2}-\frac{9\sqrt{3}}{16}k_{\theta},\\
\gamma_{0} = 0.
\end{array}
\right.
\end{eqnarray}
Inserting these equilibrium coefficients into Eq.~(\ref{eq_solution}), we obtain the angle variation without strain: $\delta_{\theta}^{+} = \frac{|\beta_{0}|}{\alpha_{0}}$ and $\delta_{\theta}^{-} = 0$. Obviously, only $\delta_{\theta}^{-}$ is the physical solution, because the angle variation should vanish without strain. For the first-order term, we have
\begin{eqnarray}
\left\{
\begin{array}{l}
\alpha_{1} = \frac{\partial\alpha}{\partial\epsilon}|_{\epsilon=0}=\frac{3}{32}k_{b}b_{0}^{2},\\
\beta_{1}  = \frac{\partial\beta}{\partial\epsilon}|_{\epsilon=0}=-\frac{\sqrt{3}}{4}k_{b}b_{0}^{2},\\
\gamma_{1} = \frac{\partial\gamma}{\partial\epsilon}|_{\epsilon=0}=\frac{3}{8}k_{b}b_{0}^{2}.
\end{array}
\right.
\end{eqnarray}
The angle variation is:
\begin{eqnarray}
\delta_{\theta} & = & \frac{\gamma_{1}}{|\beta_{0}|}\epsilon = \frac{1}{\frac{\sqrt{3}}{6}+\frac{3\sqrt{3}}{2}\eta}\epsilon,
\end{eqnarray}
where $\eta = \frac{k_{\theta}}{k_{b}b_{0}^{2}}$. According to this formula, the kink angle variation is determined by the competition between kink angle and arm length potentials. It is smaller for KSiNW with larger $k_{\theta}$, or smaller $k_{b}$, which is physically correct. A rigid arm ($k_{b}\rightarrow +\infty$) results in the largest angle variation of $\delta_{\theta}=2\sqrt{3}\epsilon$.

Inserting the kink angle variation into the total strain energy in Eq.~(\ref{eq_Etot}), and do linear approximation, we obtain the elastic strain energy density,
\begin{eqnarray*}
V_{\epsilon} & \approx & \frac{1}{Ab_{0}} \left[ \frac{3}{8}k_{\theta}\delta_{\theta}^{2}+\frac{1}{2}k_{b}b_{0}^{2}\left(\epsilon-\frac{\sqrt{3}}{6}\delta_{\theta}\right)^{2}\right]\\
 & \equiv & \frac{1}{2}Y\epsilon^{2},
\end{eqnarray*}
where $A$ is the cross section of the KSiNW. The Young's modulus is
\begin{eqnarray}
Y &=& \frac{1}{A}\times\frac{9k_{\theta}}{b_{0}+9\frac{k_{\theta}}{k_{b}}\frac{1}{b_{0}}}\nonumber\\
  &=& \frac{c_{1}}{b_{0}+c_{2}\frac{1}{b_{0}}},
\label{eq_young}
\end{eqnarray}
where coefficients $c_{1}= \frac{9k_{\theta}}{A}$ and $c_{2} = \frac{9k_{\theta}}{k_{b}}$. 

Eq.~(\ref{eq_young}) is exactly the fitting function for the MD simulation results in Fig.~\ref{fig_young}. This coincide verifies the spring representation of the KSiNW. We note that the deviation between MD simulations and the valence force field model in the figure for longer arm length is probably due to nonlinear effects resulting from larger kink angle variation for longer arm length, which is omitted in our derivation. From the above analytic derivation, it is clear that the Young's modulus is determined by the competition between the kink angle bending potential and the arm length stretching potential, corresponding to the two terms in the denominator of Eq.~(\ref{eq_young}). For KSiNW with $b_{0}< \sqrt{c_{2}}$, the Young's modulus increases with increasing arm length; while the opposite phenomenon should be observed in KSiNW with $b_{0}< \sqrt{c_{2}}$. Our MD simulation results in Fig.~\ref{fig_young} show that these KSiNW in our study fall in the latter case. The size-dependence of the Young's modulus is governed by the coefficient $c_{2}$, which is determined by the ratio of the two force constants $k_{\theta}$ and $k_{b}$. Hence, our analytic formula provides valuable information for tuning the behavior of the Young's modulus by using various materials with different $k_{\theta}$ or $k_{b}$. It is interesting that $c_{2}$ is independent of the diameter of the KSiNW, so the decreasing behavior of the Young's modulus is diameter independent, which is indeed observed in the MD simulation results.

Using the fitting parameters $(c_{1}, c_{2})$ from our MD simulations, we can extract the two force constants $(k_{b}, k_{\theta})$ to be (103.8, 265.3), (179.8, 2897.3), and (412.8, 9723.9) for KSiNWs with diameters 1.54, 3.07, and 4.61~{nm}, respectively. Dimensions of these two force constants are $[k_{b}]$=eV/\AA$^{2}$ and $[k_{\theta}]$=eV. Both force constants $k_{b}$ and $k_{\theta}$ are independent of the arm length, but increase rapidly with increasing diameter. These two force constants are effective values for the whole KSiNW system, so it can be quite different from the force constants in the potential for an actual chemical bond. For instance, $k_{\theta}$ for KSiNWs with diameter 1.54~{nm} is about 20 times larger than that describing C-C-C angles in carbon nanotubes\cite{JiangJW2008}, while $k_{b}$ is half of that in the C-C bonds in carbon nanotubes\cite{ZhouX,ChangT}. From the energy point of view, each silicon atom in the system is connected to its neighbors by small springs within the linear approximation. $k_{b}$ and $k_{\theta}$ are actually the summation over these small springs between silicon atom pairs. For KSiNW with larger diameter, there are more small springs that are in parallel connection, so the two effective force constants $k_{b}$ and $k_{\theta}$ are larger. For KSiNW with longer arm length, there is an increasing number of small springs that are in series connection, which shouldn't affect the effective force constants. As a result, $k_{b}$ and $k_{\theta}$ is sensitive to the diameter of the KSiNW, while is independent of its arm length.

\section{conclusion}
To summarize, we first point out a lattice constraint on the diameter of the KSiNW. Then we perform extensive MD simulations and finite element method to study the Young's modulus of the KSiNW. The Young's modulus is found to decrease rapidly with increasing arm length. The size sensitivity is explained by a valence force field mode, where we derived an analytic formula for the axial Young's modulus of the KSiNW with different arm length, i.e $Y=c_{1}/(b+c_{2}\frac{1}{b})$. We show that the size-dependence of the Young's modulus in KSiNW is governed by two different types of potentials stored inside its zigzag configuration, which lead to the two terms in the denominator.

\textbf{Acknowledgements} The work is supported by the Grant Research Foundation (DFG) Germany.  We thank H. S. Park at Boston University for useful discussions.

\end{document}